# A Decision Model for Supporting Task Allocation Processes in Global Software Development

Ansgar Lamersdorf[1], Jürgen Münch[2], Dieter Rombach[1,2]

[1]University of Kaiserslautern, [2]Fraunhofer IESE
a_lamers@informatik.uni-kl.de, {juergen.muench, dieter.rombach}@iese.fraunhofer.de

Today, software-intensive systems are increasingly being developed in a globally distributed way. However, besides its benefit, global development also bears a set of risks and problems. One critical factor for successful project management of distributed software development is the allocation of tasks to sites, as this is assumed to have a major influence on the benefits and risks. We introduce a model that aims at improving management processes in globally distributed projects by giving decision support for task allocation that systematically regards multiple criteria. The criteria and causal relationships were identified in a literature study and refined in a qualitative interview study. The model uses existing approaches from distributed systems and statistical modeling. The article gives an overview of the problem and related work, introduces the empirical and theoretical foundations of the model, and shows the use of the model in an example scenario.

## 1. Motivation

More and more software products are being developed in a globally distributed way: Technological advances and the possible benefits of distributed development have made this not only a common practice but also a "business necessity" ([23], [12]). The expected benefits include cost savings, access to a worldwide resource pool, proximity to customers and markets, and a reduction in overall development time through a "follow-the-sun" approach [7].

However, global software development also imposes a set of problems and risks that are often overlooked [38]: For example, communication problems, caused by distance, language, and cultural differences, reduce productivity ([21], [22]) and quality suffers from inexperienced developers at remote sites or from a lack of trust between distributed teams [41]. These problems can even annihilate the cost reduction of sending work to low-cost regions [38].

In order to address the benefits and, at the same time, the risks and problems of global software development, effective project management is needed that actively considers the nature and characteristics of global software development. An important activity in global software development project management is task allocation: In addition to having to consider the characteristics and the availability of the workforce (as in collocated development), task allocation in global software development must take into account the characteristics of the sites and their relationships (such as time zone differences or infrastructure).

Depending on the focus and the goals of a software development project, different allocations might be suited differently: In order to increase productivity, independent chunks of work should be assigned to every site [32]. On the other hand, assigning interdependent tasks to sites in different time zones might decrease the development time [7]. The lowest labor rates can be achieved by assigning as much work as possible to low-cost sites.

These goals and assignment strategies sometimes conflict with each other and have to be regarded systematically in order to identify the best task allocation for a particular project. In practice, however, allocation is not done systematically and often considers only single aspects such as labor costs [3]. Thus, there is a need for improving management processes in globally distributed software development processes.

This article presents a method for improving task allocation processes by developing a model for decision support. The model uses multiple criteria and weighted goals as input for suggesting a weighted list of possible task assignments. It is based on a systematic literature review and an interview study conducted in order to identify the factors that influence the success of distributed development projects.

The remainder of this article is structured as follows: Section 2 gives an overview of the related work in models for task allocation. The model is presented in detail in Section 3 together with its goals, a systematic literature review for determining its criteria and causal relationships, and a demonstration of its use within an example project. Section 4 names the limitations of the model and Section 5 concludes the article.

## 2. Related Work

In [32], a simple model for task allocation in global software development is presented. The underlying assumption is that software development can be described as a series of modification requests to a set of modules. Based on that, an algorithm is developed, which, for a given set of modules and modification requests, tries to find the optimal assignment of modules to sites. Optimal here means that the number of modification requests spanning multiple sites is minimized in order to reduce communication overhead.

The model represents a formal and well-defined approach for optimizing task allocation. However, its main drawback is the fact that it only considers one single criterion, namely, minimization of the communication needed between the available sites. It also uses the available resources per site as a constraint, but essential factors that influence project success (e.g., the available expertise or the cost rate per site) are not considered.

Another model for task allocation was developed by Setamanit, Wakeland, and Raffo [39]. Based on a combination of discrete-event and system-dynamic simulation, it allows for evaluating different allocation strategies. The model simulates software development at every site as well as the effects of the interaction between sites. Thus, it is able to make statements on the effects of different strategies on productivity.

However, the sites are only rudimentarily described in the model. Therefore, the model can only make general statements and cannot be used for concrete decision support. Besides, the factors influencing productivity are not identified empirically; thus, it remains unclear if they truly reflect the factors relevant in practice.

Other models for assigning tasks to a set of sites exist in other domains: In production, algorithms have been developed for allocating production work to a network of global sites with the goal of minimizing production and transportation costs. In the distributed systems domain, there are approaches for optimizing the allocation of computing tasks to a set of processors. An analysis and comparison of existing approaches was done in [30]. The approaches were evaluated against a set of requirements for a task allocation model in GSD. The result showed that none of the models fulfilled all requirements.

However, one algorithm for task allocation in distributed systems by Bokhari [6] satisfied most of the requirements compared to the other approaches. The algorithm tries to minimize the sum of the execution costs of the tasks at the processors and the costs of transmitting data between tasks at different processors. The main drawbacks for its application in GSD are: 1) The algorithm obviously does not contain empirical data on distributed development. Particularly, it does not contain a set of variables that represent the relevant characteristics of GSD. 2) The algorithm needs exact numbers as input. For example, the cost of processing a specific task at a specific processor has to be described with an exact number. Such a number can often not be specified when human behavior is modeled.

In the following, a model is proposed that reuses the algorithm while also addressing these drawbacks.

## 3. The Decision Model

The following section will introduce the decision model. First, the terminology and model goals are given. The model is based on a combined literature review and interview study on the criteria and causal relationships in task assignment that will be shown second. Afterwards, the theoretical foundations of the model will be presented, followed by the application of the model in an example project.

### 3.1. Terminology and Model Goals

The underlying assumption of the model is that every software development project consists of a weighted set of *goals* that define project success (e.g., project costs, software quality).

Project management in global software development aims at fulfilling these goals by assigning the tasks of a software development project to the appropriate sites during *task assignment*.

However, the effect of the task assignment on project goals depends on a set of *characteristics* of distributed software development. Time shift between sites, for example, is such a characteristic: If tasks are assigned to two sites with a large time shift between them, productivity may be reduced and thus project costs would in-crease. Task assignment should thus not only consider the project goals but also the characteristics of distributed development.

Project goals and characteristics of distributed development together represent the criteria that should be regarded in task assignment for global software development. This is the main goal of the decision support model presented here that considers these criteria.

More formally, the main goal can be described as follows: From the perspective of a project manager in a global software development project, it is the purpose of the model to support task allocation with respect to individual project goals and characteristics of distributed development.

From that goal, the following sub-goals are derived:
- Task allocation should be supported by suggesting several assignments of tasks to sites for a given project. Using these suggestions, the project manager can then make improved, systematic allocation decisions.
- The model should consider individual project goals. Therefore, the suggestions made by the model should be dependent on the priorities of the project.
- The characteristics of globally distributed development (e.g., the overhead of working and communicating in a distributed manner) should be taken into account systematically.

Further, more detailed, requirements for a decision support model are defined in [30]: A distribution model should support *multiple goals*, should be able to describe both *properties of tasks and sites* and *dependencies between tasks and sites*, and should be *adaptable* to different environments. An appropriate degree of *formality* should allow for making suggestions automatically and the criteria and causal relationships used in the model should be *empirically based*.

### 3.2. Empirical Identification of Criteria and Causal Relations

The empirical foundations of the model were laid using a combined literature review and interview study on distributed software development. These resulted in a set of criteria and causal relationships. The results were then used for the development of the task allocation model. The study is summarized in the following. (It is explained in more detail in [29])

The goal of the literature and interview study can be described as follows: From the perspective of a project manager in a global software development project, the criteria for task assignment and the underlying causal relationships should be identified. Three research questions were derived from that:
- Question 1: What are the goals of distributed development projects?
- Question 2: What characteristics of distributed development should be regarded during task assignment?
- Question 3: What are the relationships between the characteristics of distributed development and project goals?

The following steps were performed in the study:

1. Literature study: A literature study was conducted first. 26 publications from different journals, conferences, and workshops were analyzed. They can be classified into case studies, empirical studies (reporting the experiences of several distributed development projects), and other types of publications. Table 1 lists the analyzed literature. As a result, a first set of criteria and causal relationships was identified.

**Table 1.** Analyzed literature

| Case Studies | Empirical Studies | Other |
|---|---|---|
| [42], [4], [14], [33], [31], [9], [20], [26] | [2], [34], [28], [15], [24], [25], [41], [35], [36], [18], [40], [27], [16], [13], [10] | [37], [19] [8] |

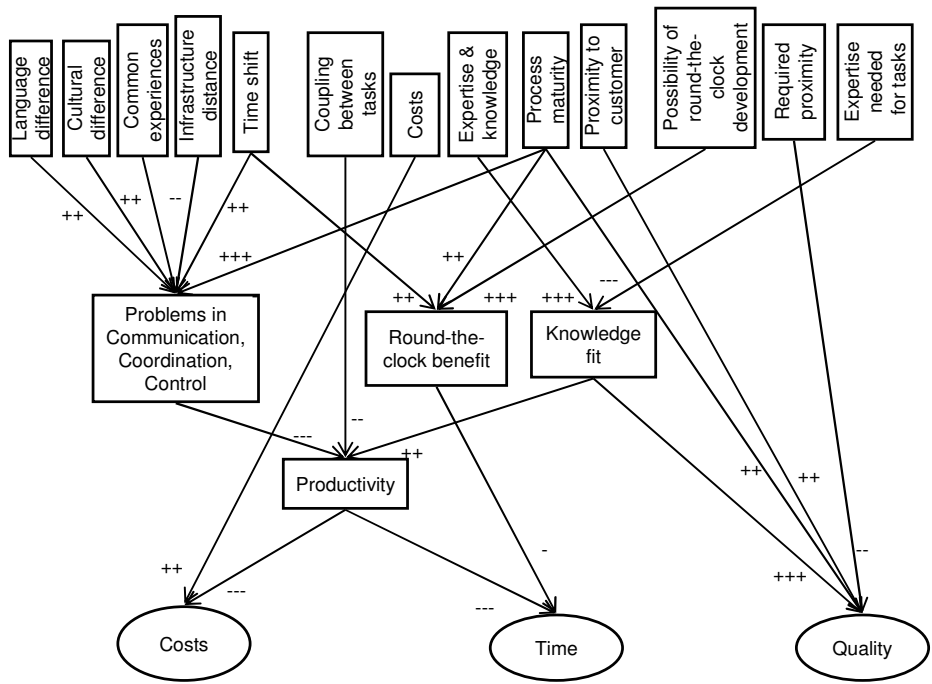

**Fig. 1.** Identified goals, influencing factors, and their relationship: strong (+++), medium (++), or soft (+) impact that is positive (+) or negative (-)

2. Questionnaire design: Based on the literature results, a questionnaire was designed for use in interviews with practitioners. In the questionnaire, the findings from the literature study were presented and the practitioners were asked to comment on these results.

3. Interview study: An interview study was conducted with managers of distributed software development. Interviews were conducted either in person or over the telephone. They usually lasted for approximately one hour. The interviews were part of a larger study on distributed development (see [29]), with ten of them being used for the work presented here. All interviews were recorded and transcribed literally.

4. Analysis: The transcribed interviews were analyzed question by question, comparing the answers with the literature study results. According to the practitioners' answers, the previous findings were weighted, new criteria and causal relationships were added, and irrelevant factors were removed.

The study resulted in a set of 13 influencing factors. These factors have an influence on four intermediate factors (problems in communication, coordination, and control; possible benefit of round-the-clock-development; productivity; fit between the knowledge needed for a task and that available at a site) and on three goals (cost, time, quality). Figure 1 shows the relationships identified between influencing factors and goals. It also gives a relative weight for the (positive or negative) influences.

### 3.3. Model Overview

Based on the results of the literature and interview studies, a model for supporting task allocation decisions was developed. The algorithms of the model reuse approaches from distributed systems and statistical modeling. In this section, the main elements and algorithms of the model are sketched.

#### 3.3.1. Distributed Systems Algorithm for Identifying Optimal Assignments

In an earlier study [30], the distributed systems algorithm of Bokhari was identified as most promising for reuse in a GSD distribution model. A detailed explanation of the model can be found in [6].

The algorithm gets as input a set of modules (i.e., tasks) and a set of processors the modules can be assigned to. It considers two kinds of costs:

- Costs of executing module $i$ on processor $p$. These are described as $e_{ip}$.
- Costs of transmitting data between module $i$ and module $j$ with $i$ being assigned to processor $p$ and $j$ to $q$. These are described as $s_{pq}(d_{ij})$ with $d_{ij}$ representing the amount of data transmitted between modules $i$ and $j$ and $s_{pq}$ being the cost for transmitting one unit of data between $p$ and $q$.

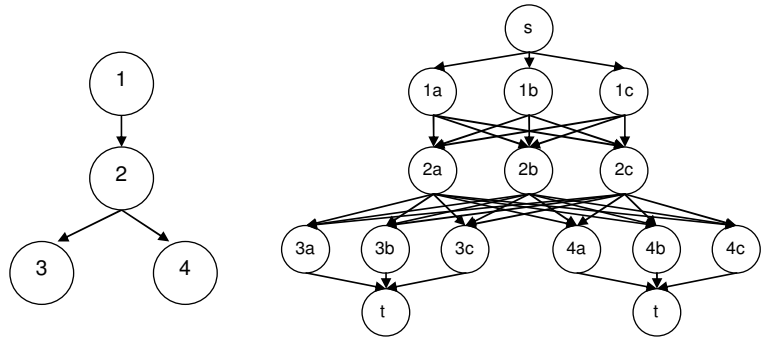

**Fig. 2.** An invocation tree and the corresponding assignment graph for three processors (a, b, c)

The tasks are assumed to be connected in a tree structure – every module is called by a single parent module and can call a set of other modules. This structure is called an *invocation tree*. The algorithm creates an *assignment graph* out of the invocation tree by creating a node for every combination of module and processor and connecting them in accordance with the invocation tree (see Figure 2).

The edges in the assignment graph are weighted with the combined execution and transmission costs. A graph algorithm developed by Bokhari then uses a dynamic programming approach for efficiently identifying the shortest paths through the graph. These paths represent the optimal assignment of modules to processors with a minimal sum of all execution and transmission costs.

On a high-level view, the algorithm solves a problem similar to the task assignment in GSD. Applying the model to GSD means:
- Modules and processors are represented by tasks and sites.
- The costs of executing module *i* on processor *p* are represented by the effort of doing task *i* of a software development project at site *p* (mainly depending on the characteristics of tasks and sites identified in Section 3.2).
- The costs of transmitting data between module *i* and module *j* with *i* being assigned to processor *p* and *j* to *q* are represented by the overhead being created between tasks *i* and *j* that are assigned to sites *p* and *q* (mainly depending on the dependencies between tasks and sites identified in Section 3.2).

The input variables describing the cost functions $e_{ip}$ and $s_{pq}(d_{ij})$ in detail are given by the results of the empirical study. However, other problems remain:
- The algorithm can only handle tasks that are connected in a tree structure. However, tasks in a development project can have arbitrary connections.
- Costs are the only criteria for comparing different assignments. Therefore, the different conflicting goals that can exist in global software development have to be aggregated into one cost function.
- All costs are described by a single, distinct number, which does not represent the reality of human development that contains a large amount of uncertainty.

The first problem was solved by developing an extension of Bokhari's algorithm that contains an additional first step of transferring arbitrary graphs into a set of trees (however, with reduced efficiency). The other two problems were solved by describing the cost functions not by single numbers but using Bayesian networks.

### 3.3.2. Bayesian Networks for Evaluating Assignments

A Bayesian Network (BN) is able to formulate causal relationships under conditions of uncertainty. It consists of a directed acyclic graph representing discrete variables and their relationships and a set of probability tables. For every variable, one table describes the probabilities of its values as a function of the input variables [5].

The application of mathematical methods allows for inference within BNs: Using bottom-up and top-down reasoning, statements can be made on the probabilistic distribution of the values of any variable based on a set of observed values of other variables. In addition, it is possible to make reasoned statements even if not all independent variables have defined observed values. Thus, in software engineering research, BNs have been used to model and predict software development projects [17].

We used Bayesian networks in our model to represent the cost functions of the distributed systems algorithm of Bokhari: Both the cost of executing a task at a site and the cost of transmitting data between sites is represented by a BN. Figures 3 and 4 show the resulting networks.

Every BN models the impact of a set of input variables on three cost types (financial, time, quality). This is done for every combination of task and sites individually. For example, the BN for describing the cost at a site (Figure 3) can be instantiated for task $t_1$ and site $s_1$ with the according parameters of $t_1$ and $s_1$ (e.g., the size of $t_1$ and the process maturity at $s_1$).

BNs operate with discrete values for every input and output variable. We thus defined five steps from "very low" to "very high" for most variables (e.g., proximity to customer). For other variables (e.g., cost rate) that have numeric values, we defined intervals in order to get discrete values.

The probabilistic tables for the BNs were designed with help of the AgenaRisk tool [1]. It contains functions for calculating the table values by using the normal distribution and by representing the discrete values with numbers from 1 to 5. For example, the table for "development quality" is calculated by generating a normal distribution with the weighted average of "staff capability" and "process maturity" as mean value. The integration of this function between the intervals $(0, 1)… (4, 5)$ then delivers the values for the probabilistic table.

Input variables, cost variables, causal relationships, and their weights (e.g., the weights of "staff capability" and "process maturity" on "development quality") were taken from the results of the literature and interview studies.

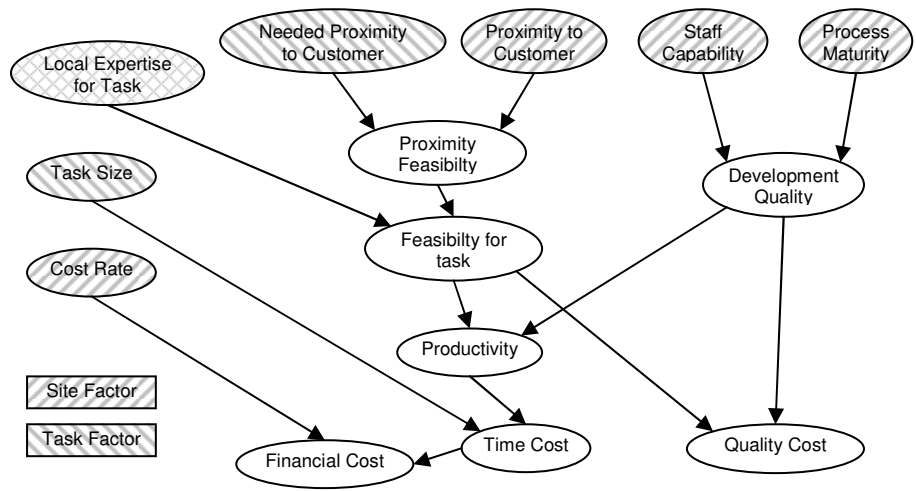

**Fig. 3.** Bayesian network for cost at site

In order to get one single cost function, all three costs (financial, time, quality) are normalized and added with different weights (which are dependent on project priorities) into one function.

The repeated application of the two networks for every combination of tasks and sites makes it possible to describe the needed cost functions of the distributed systems algorithm. However, the values of the functions are not distinct numbers but probabil-istic distributions over a set of cost values. This makes the uncertainty in human behavior explicit. On the other hand, Bokhari's algorithm uses distinct values as input. Therefore, an algorithm was developed that is able to suggest assignments by using the distributed systems algorithm while taking the probabilistic cost distributions as input.

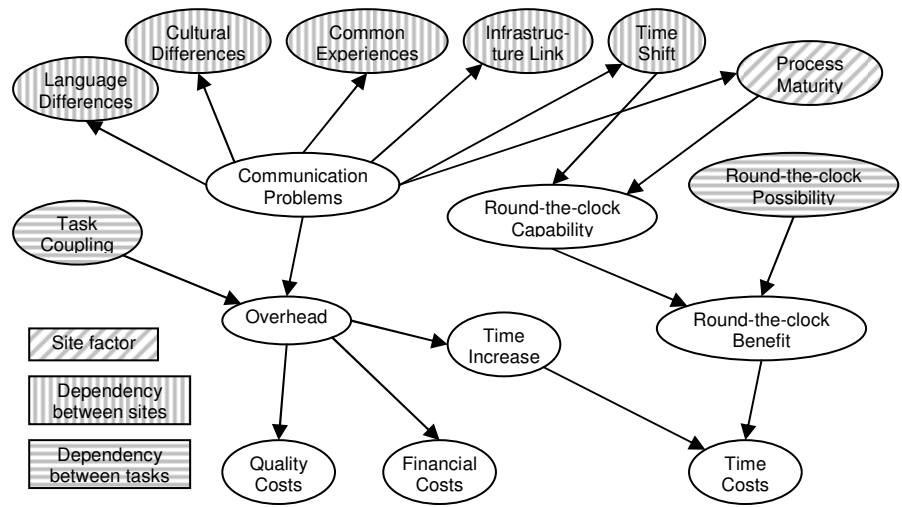

**Fig. 4.** Bayesian network for transmission cost

### 3.3.3 Algorithm for Suggesting Assignments
The link between the Bayesian networks results and the (adapted) algorithm of Bokhari is provided by a randomization algorithm. It basically consists of three steps:
- Collect the probabilistic distributions by executing the BNs for every combination of tasks and sites.
- Repeat for a large number of runs:
    o Randomly pick one number out of every probabilistic distribution. The probabilities for every random pick are provided by the probabilistic distributions. Store the numbers as cost functions for the distributed systems algorithm.
    o Execute the distributed system algorithm and store the returned assignment.
- Return the stored assignments in an ordered list with a decreasing number of occurrences.

In other words, the algorithm simulates a number of scenarios with randomly chosen numbers for the individual cost functions, based on the probabilistic distributions. This ensures, on the one hand, that across all scenarios, the costs reflect the predictions of the Bayesian networks. On the other hand, within each run, all costs are represented by distinct numbers, which makes the execution of Bokhari's algorithm possible.

As a result, the algorithm returns not one but several ordered assignments together with information on the number of scenarios in which each distribution was optimal. This makes the uncertainty in predicting human behavior explicit and gives the project manager the opportunity to choose from an ordered set of assignments.

### 3.4. Example

The model was implemented as a Java prototype with a Swing GUI and consisted of a generic and a model-specific part. The generic part contained implementations of the algorithm of Bokhari, the randomization algorithm, and the Bayesian networks. The BN implementation reused the JavaBayes framework [11] and extended it with functions for calculating the probabilistic tables similar to the functions used in the AgenaRisk [1] tool. The model-specific part implemented the BNs that were derived from the empirical study. As these were developed using AgenaRisk, they were transformed by hand into the implementation.

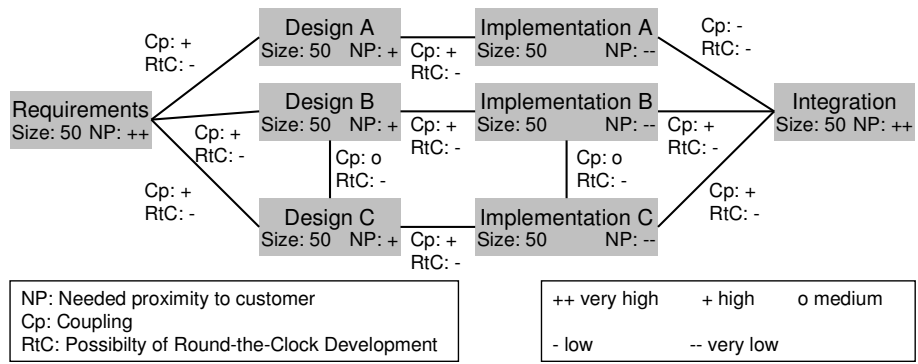

**Fig. 5.** Project example – Tasks to be distributed

In the following, the use of the model will be shown in a hypothetical example. The tasks of the example project include requirements engineering, design and implementation of three different components, and integration. Three sites are available: One site at the customer, which is very expensive but has very good skills in requirements engineering and design. The second site is in the US. It is also expensive (but not as much as the customer's site) and also has good skills in requirements engineering and design. The Indian site has large differences (especially in language and culture) compared to the other two sites, but is very inexpensive. People there have very good skills in implementation but are inexperienced in requirements engineering and design. Figures 5 and 6 show the tasks and sites with their parameters in detail.

Table 2 shows the results of executing the model with three different weights on the goals. For every execution the three best results are presented together with the number of runs the assignment was optimal (e.g., in the first execution, the best assignment was optimal in 9% of the runs). In the first result, the focus was on all goals, with the highest weight on quality (Cost: 20%, Time: 30%, Quality: 50%). Here the model suggests doing the implementation in India and requirements and designing either at the customer's site or at the US site. Integration should be done at the customer's site (because it should be close to the customer) or in Asia (because it is closely coupled with implementation).

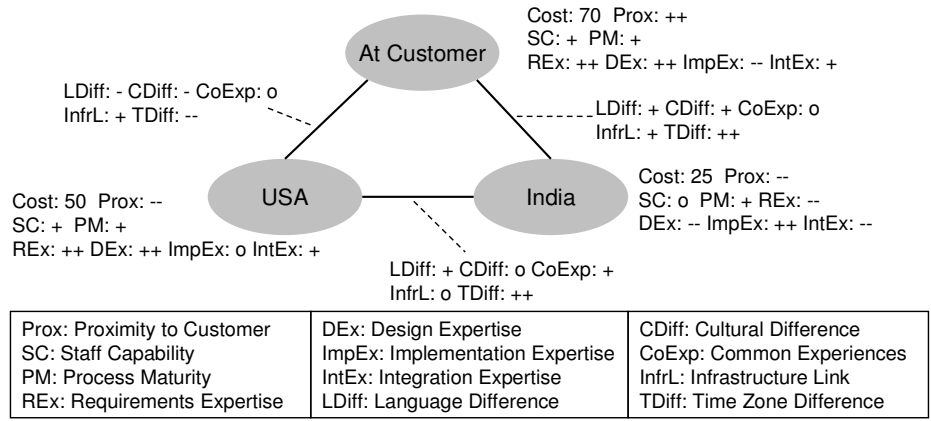

**Fig. 6.** Available sites

The next result shows the execution of the model with a very strong focus on the costs and very little regard for time and quality (Cost: 80%, Time: 10%, Quality: 10%). It can be seen that the model then suggests doing everything in India due to the low cost rate there. An alternative would be assigning requirements and design to the US site.

In the last run, the focus was set primarily on development time (Cost: 10%, Time: 80%, Quality: 10%). Now, the model favors assigning all tasks to one site, since this would reduce the overhead of distributed communication. Another alternative given by the model is to do every task at the site that has the best knowledge, which means assigning implementation to Asia and requirements and design to the customer site.

## 4. Limitations and Validity of the Model

There are several limitations regarding the applicability of the model:

The experiences gathered in the empirical study come from many different organizations and project environment. Therefore, the expressed relationships describe a general overview rather than a concrete environment. Within a specific organization, the relative weights of the criteria may differ, or additional criteria may be relevant. The model thus has to be adapted in order to be used in a specific environment. However, due to its modularization, this can be done by changing the Bayesian networks without having to modify the algorithms.

Underlying the model development was the assumption that project management can divide a project upfront into distinct tasks that can be independently assigned to the available sites. However, a project manager often has no clear information on the tasks of a project because, for example, an agile process is followed or there is not enough knowledge on the requirements or the technology. In these cases, it would be hard to use the model. This also implies that the model evaluation should start using historic project data as it is easier to identify distinct tasks in retrospective.

**Table 2.** Model results with focus on quality (left), development costs (middle), and development time (right)

| 1.:9% | | Cust | US | Asia |
|---|---|---|---|---|
| | Reqs | X | | |
| | Des A | X | | |
| | Impl A | | | X |
| | Des B | X | | |
| | Impl B | | | X |
| | Des C | | X | |
| | Imp C | | | X |
| | Integr | X | | |

| 1.:51% | | Cust | US | Asia |
|---|---|---|---|---|
| | Reqs | | | X |
| | Des A | | | X |
| | Impl A | | | X |
| | Des B | | | X |
| | Impl B | | | X |
| | Des C | | | X |
| | Imp C | | | X |
| | Integr | | | X |

| 1.:18% | | Cust | US | Asia |
|---|---|---|---|---|
| | Reqs | X | | |
| | Des A | X | | |
| | Impl A | X | | |
| | Des B | X | | |
| | Impl B | X | | |
| | Des C | X | | |
| | Imp C | X | | |
| | Integr | X | | |

| 2.:8% | | Cust | US | Asia |
|---|---|---|---|---|
| | Reqs | | X | |
| | Des A | | X | |
| | Impl A | | | X |
| | Des B | | X | |
| | Impl B | | | X |
| | Des C | | X | |
| | Imp C | | | X |
| | Integr | X | | |

| 2.:7% | | Cust | US | Asia |
|---|---|---|---|---|
| | Reqs | | X | |
| | Des A | | X | |
| | Impl A | | | X |
| | Des B | | X | |
| | Impl B | | | X |
| | Des C | | X | |
| | Imp C | | | X |
| | Integr | | | X |

| 2.:11% | | Cust | US | Asia |
|---|---|---|---|---|
| | Reqs | X | | |
| | Des A | X | | |
| | Impl A | | | X |
| | Des B | X | | |
| | Impl B | | | X |
| | Des C | X | | |
| | Imp C | | | X |
| | Integr | | | X |

| 3.:7% | | Cust | US | Asia |
|---|---|---|---|---|
| | Reqs | X | | |
| | Des A | X | | |
| | Impl A | | | X |
| | Des B | X | | |
| | Impl B | | | X |
| | Des C | | X | |
| | Imp C | | | X |
| | Integr | | | X |

| 3.:6% | | Cust | US | Asia |
|---|---|---|---|---|
| | Reqs | | | X |
| | Des A | | X | |
| | Impl A | | | X |
| | Des B | | | X |
| | Impl B | | | X |
| | Des C | | | X |
| | Imp C | | | X |
| | Integr | | | X |

| 3.:10% | | Cust | US | Asia |
|---|---|---|---|---|
| | Reqs | | X | |
| | Des A | | X | |
| | Impl A | | X | |
| | Des B | | X | |
| | Impl B | | X | |
| | Des C | | X | |
| | Imp C | | X | |
| | Integr | | X | |

The model also assumes that there is enough knowledge in an organization for describing the characteristics of the sites (e.g., knowledge available, cultural differences). In Bayesian networks, it is possible to calculate probabilistic distributions without all input parameters having distinct values. Therefore, the model can be used even if not all variables are known. But the less information is known, the less useful are the suggestions made by the model.

The BNs operate with variable values from "very low" to "very high". As they are relatively fuzzy and subjective, an application of the model in a real-world environment needs to come with specific evaluation guidelines (e.g., which time zone distance is to be interpreted as "low" and which as "medium").

Although the criteria and causal relationships of the model presented here stem from an empirical study, the model needs further evaluation. It has so far only been used for simulating task assignment processes with hypothetical input data. Therefore, external validity needs to be carefully considered when applying the model and making conclusions in practice.

## 5. Conclusion and Future Work

The main goal of the work presented here was to find decision support for task allocation that considers multiple criteria for the decision. It is, however, not easy to clearly define the term "criteria" in a conceptual framework for a model. We distinguished between *goals* of software development projects (cost, time, quality) and *characteristics of distributed development* that have an impact on the goals. Based on that assumption and on an empirical study, we developed a model for decision support in task allocation that reuses an approach from distributed systems and Bayesian networks in order to suggest a prioritized list of assignments.

By conducting an empirical study on the goals and characteristics of distributed development, we assured that the model considered criteria relevant for task allocation. However, since the adapted distributed systems algorithm and the mechanism of selecting cost values according to probabilistic distributions work independently of the Bayesian networks, the model can be easily changed if other goals or influencing factors are relevant in a specific environment.

The model fulfills the goals stated in Section 3.1.: It results in a weighted list of *suggestions for task allocation* while systematically considering both *multiple project goals* and *characteristics of distributed development*. The requirements for a distribution model defined in [30] are also fulfilled:

- Multi-objectivity: The example shows how different weights put on the project goals can change the resulting assignments suggested by the model.
- Properties of tasks and sites, dependencies between tasks and sites: All of these types of influencing factors can be described in the Bayesian networks.
- Adaptability: The model can be adapted to different environments by changing the Bayesian networks.
- Formality: The model contains formal algorithms that can automatically suggest assignments.
- Empirically-based criteria: The influencing factors and goals were identified in an empirical study.

Future work will have to test the model in real-world environments. We therefore plan to evaluate and iteratively extend the model in case studies and experiments.